\documentclass[11pt]{article}
\usepackage[margin=1in]{geometry}
\usepackage{graphicx}
\usepackage{amsmath, amssymb}
\usepackage{booktabs}
\usepackage{siunitx}
\usepackage{hyperref}
\usepackage{caption}
\usepackage{subcaption}
\usepackage{multirow}
\usepackage{url}
\usepackage{enumitem}
\setlist{nosep}
\usepackage{float}     
\usepackage{placeins}  
\usepackage{xcolor}

\hypersetup{
  colorlinks=true,
  linkcolor=blue,
  citecolor=blue,
  urlcolor=blue
}

\title{\bf An Explainable AI Approach for Lung Cancer Detection Using\\ Convolutional Neural Networks}
\author{
  Nishan Rai$^{1}$\thanks{Email: nishanrai.075@kathford.edu.np} \and
  Sujan Khatri$^{1}$ \and
  Devendra Risal$^{1}$ \\
  $^{1}$Kathford International College of Engineering and Management, Lalitpur, Nepal
}
\date{2022}

\begin{document}
\maketitle

\begin{abstract}
\noindent
Early detection of lung cancer is critical to improving survival outcomes. We present a deep learning framework for automated lung cancer screening from chest computed tomography (CT) images with integrated explainability. Using the IQ{-}OTH/NCCD dataset (1{,}197 scans across Normal, Benign, and Malignant classes), we evaluate a custom convolutional neural network (CNN) and three fine{-}tuned transfer learning backbones: DenseNet121, ResNet152, and VGG19. Models are trained with cost-sensitive learning to mitigate class imbalance and are evaluated using accuracy, precision, recall, F1-score, and ROC-AUC. While ResNet152 achieved the highest accuracy (\SI{97.3}{\percent}), DenseNet121, although having slightly lower accuracy than the custom CNN, achieved the most balanced precision and recall, which is critical in medical screening. We further apply Shapley Additive Explanations (SHAP) to visualize evidence contributing to predictions, improving clinical transparency. Results indicate that CNN-based approaches augmented with explainability can provide fast, accurate, and interpretable support for lung cancer screening, particularly in resource-limited settings.
\end{abstract}

\section{Introduction}
Lung cancer is one of the leading causes of cancer mortality worldwide. Early diagnosis substantially improves survival, yet accurate reading of CT scans is time-consuming and depends on radiologist expertise and availability. In low-resource or rural settings, the shortage of trained radiologists further delays screening and follow-up. Deep learning, particularly convolutional neural networks (CNNs), has demonstrated strong performance in visual recognition tasks and is increasingly used in medical imaging. However, the adoption of AI in clinical workflows requires not only high performance but also transparency and interpretability to support safe decision-making.

This work develops and evaluates an explainable AI (XAI) pipeline for lung cancer detection from CT images. We compare a custom CNN against three widely used transfer learning backbones---DenseNet121, ResNet152, and VGG19---and integrate SHAP to visualize model evidence. Our contributions are:
\begin{itemize}
  \item A complete training and evaluation pipeline on a public, multi-class CT dataset (Normal/Benign/Malignant) with data augmentation and class-imbalance handling.
  \item A comparative study of custom vs.\ fine-tuned CNNs for lung cancer detection, including DenseNet121, ResNet152, and VGG19.
  \item An explainability layer using SHAP to highlight image regions driving predictions for clinical interpretability.
\end{itemize}

\section{Related Work}
Early lung cancer detection in CT imaging has been studied extensively across traditional machine learning and modern deep learning pipelines. 

\subsection{Traditional Machine Learning Approaches}
Prior to the widespread use of CNNs, lung nodule detection relied on handcrafted feature extraction combined with classifiers such as Support Vector Machines (SVM)~\cite{valente2016svm}, k-Nearest Neighbors, and Random Forests~\cite{gonzalez2016rf}. Features often included shape, texture, and intensity statistics, sometimes extracted after segmentation using thresholding or active contour methods. While interpretable, these approaches suffered from limited generalization due to hand-crafted feature design and variability in CT acquisition protocols.

\subsection{Deep Learning with CNNs}
The introduction of deep CNNs such as AlexNet~\cite{krizhevsky2012alexnet}, VGG~\cite{simonyan2015vgg}, ResNet~\cite{he2016resnet}, and DenseNet~\cite{huang2017densenet} transformed medical image analysis by enabling end-to-end feature learning directly from pixel data. Notable works have applied 2D CNNs to slice-based classification~\cite{shen2017multicnn} and 3D CNNs for volumetric nodule analysis~\cite{zhu20183dcnn}. Transfer learning from ImageNet-pretrained models has been shown to accelerate convergence and improve performance in small medical datasets~\cite{tajbakhsh2016transfer}. Public datasets such as LIDC-IDRI~\cite{armato2011lidc} and Kaggle Data Science Bowl have served as benchmarks for these methods.

\subsection{Advanced Architectures and Attention Mechanisms}
More recent works integrate attention modules~\cite{wang2017residualattention} or Vision Transformers (ViT)~\cite{dosovitskiy2020vit} to enhance global context modeling, which is especially relevant for detecting subtle malignant features across slices. Multi-scale and multi-view CNN frameworks~\cite{shen2015multiscale} combine features from different resolutions or anatomical planes to improve robustness.

\subsection{Explainable AI in Medical Imaging}
Interpretability remains a key requirement for AI adoption in healthcare. Gradient-based saliency methods such as Grad-CAM~\cite{selvaraju2017gradcam} and Integrated Gradients~\cite{sundararajan2017integrated} provide visual explanations but can be noisy. Model-agnostic methods like LIME~\cite{ribeiro2016lime} and SHAP~\cite{lundberg2017shap} offer more stable attributions and can be applied to any classifier. Several works~\cite{holzinger2019xai} emphasize that explainable AI can help bridge the trust gap between AI predictions and clinician acceptance.

\subsection{Our Contribution in Context}
Compared to prior studies, our work uniquely combines (i) multiple CNN backbones including a custom architecture and fine-tuned DenseNet, ResNet, and VGG; (ii) cost-sensitive learning to address class imbalance; and (iii) SHAP-based interpretability specifically tailored to multi-class lung cancer classification in CT scans.

\section{Materials and Methods}
\subsection{Dataset}
\begin{figure}[h]
  \centering
  \includegraphics[width=\linewidth]{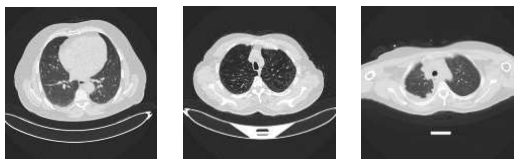}
  \caption{Representative CT samples for (left to right): Benign, Normal, Malignant.}
  \label{fig:dataset_samples}
\end{figure}

\begin{figure}[h]
  \centering
  \includegraphics[width=0.75\linewidth]{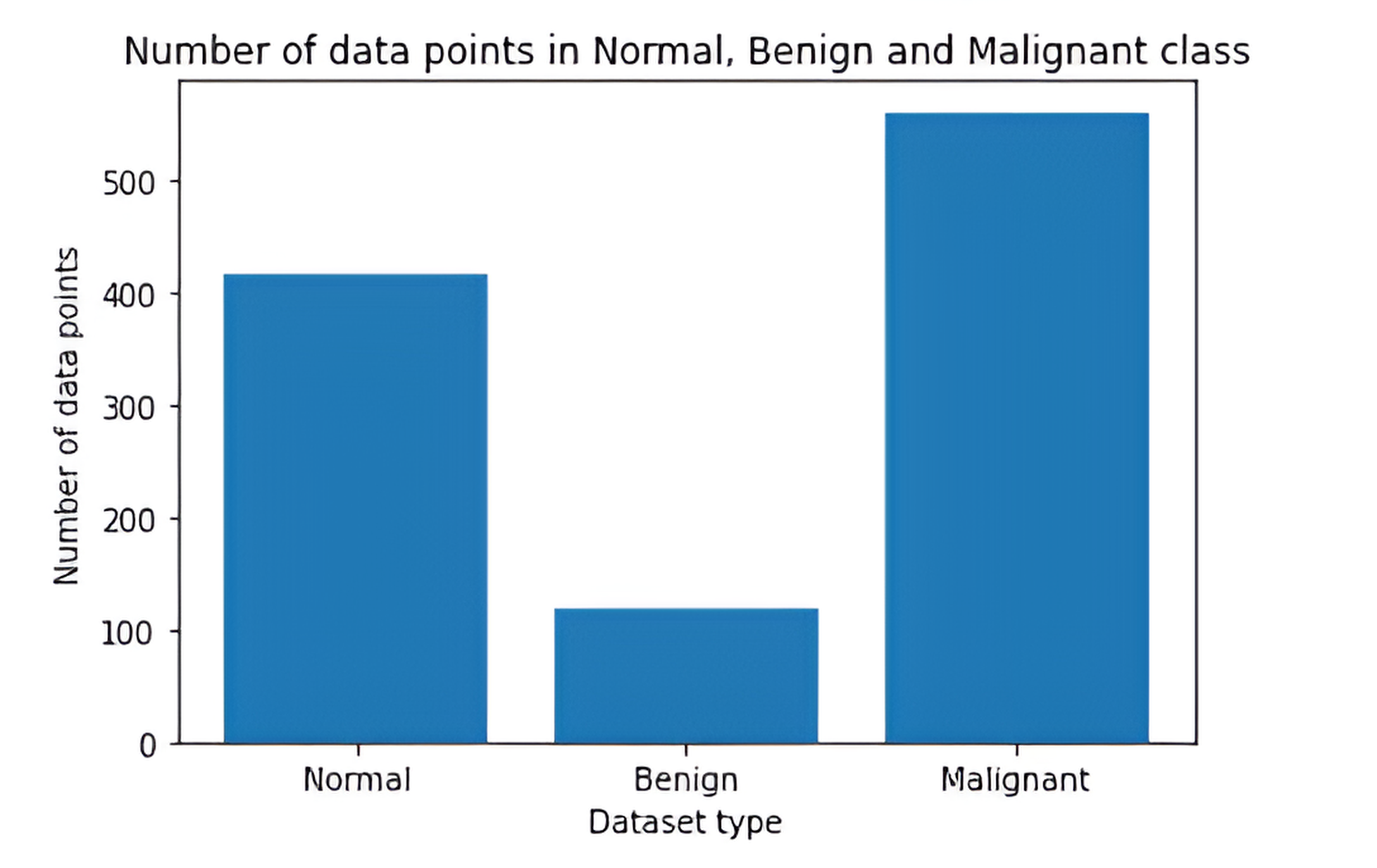}
  \caption{Class distribution in IQ{-}OTH/NCCD dataset.}
  \label{fig:class_dist}
\end{figure}

We use the IQ{-}OTH/NCCD dataset~\cite{iqoth_dataset}, comprising \num{1197} CT images labeled as Normal (416), Benign (120), and Malignant (561). Figure~\ref{fig:dataset_samples} shows representative samples; Figure~\ref{fig:class_dist} illustrates class distribution.

\subsection{Preprocessing and Augmentation}
All images are resized to $256{\times}256$. Data augmentation includes random rotations (up to $\pm 15^\circ$), center cropping, and random horizontal flips. Images are normalized using ImageNet statistics. To address class imbalance, we employ cost-sensitive learning by weighting the loss to upweight minority classes.

\subsection{Model Architectures}
\paragraph{Custom CNN.}
Our custom CNN stacks convolutional blocks (Conv--BN--ReLU--MaxPool), followed by a flattened representation and fully connected layers with dropout. Table~\ref{tab:custom_arch} summarizes the key layers and parameter counts.

\begin{table}[h]
  \centering
  \caption{Custom CNN architecture summary.}
  \label{tab:custom_arch}
  \begin{tabular}{@{}llr@{}}
    \toprule
    Layer & Output Shape & Params \\
    \midrule
    Conv(3$\to$6), BN, ReLU, MaxPool & $6 \times 256\times256 \to 6 \times 128\times128$ & 168 \\
    Conv(6$\to$12), BN, ReLU, MaxPool & $12 \times 128\times128 \to 12 \times 64\times64$ & 660 \\
    Conv(12$\to$32), BN, ReLU, MaxPool & $32 \times 64\times64 \to 32 \times 32\times32$ & 3{,}488 \\
    Flatten \& FC(32768$\to$5461) + Dropout & $-$ & 178{,}951{,}509 \\
    FC(5461$\to$3) & $-$ & 16{,}386 \\
    \bottomrule
  \end{tabular}
\end{table}

\paragraph{Transfer Learning Backbones.}
We fine-tune ImageNet-initialized DenseNet121~\cite{huang2017densenet}, ResNet152~\cite{he2016resnet}, and VGG19~\cite{simonyan2015vgg} by replacing and training the classifier head (with dropout for regularization). Trainable parameter counts after freezing the feature extractor are listed in Table~\ref{tab:tl_params}.

\begin{table}[h]
  \centering
  \caption{Trainable parameters after fine-tuning (classifier head).}
  \label{tab:tl_params}
  \begin{tabular}{@{}lr@{}}
    \toprule
    Backbone & Trainable Params \\
    \midrule
    DenseNet121 & 144{,}667 \\
    ResNet152 & 525{,}315 \\
    VGG19 & 2{,}308{,}711 \\
    \bottomrule
  \end{tabular}
\end{table}

\subsection{Training Protocol}
Models are trained using Adam optimizer with learning rate $1\times10^{-2}$, batch size 8, for up to 50 epochs, using stratified splits for train/validation/test. We monitor validation loss for early stopping and select the checkpoint with the best validation performance.

\subsection{Evaluation Metrics}
We report overall Accuracy, class-averaged Precision, Recall, and F1-score, and plot ROC curves with AUC for each class. Confusion matrices provide per-class error structures.

\subsection{Explainability via SHAP}
We apply SHAP~\cite{lundberg2017shap} to attribute model predictions to input regions. We visualize positive (evidence-for) and negative (evidence-against) contributions for Normal, Benign, and Malignant predictions, supporting clinician validation.

\section{Results}
\FloatBarrier
\subsection{Learning Curves}
\begin{figure}[H]
  \centering
  \includegraphics[width=\linewidth]{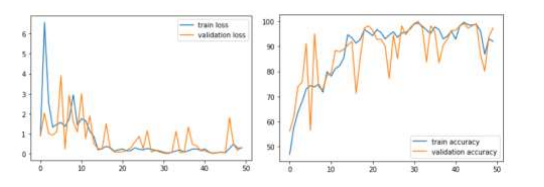}
  \caption{Learning curves for Custom CNN.}
  \label{fig:learn_custom}
\end{figure}

\begin{figure}[H]
  \centering
  \includegraphics[width=\linewidth]{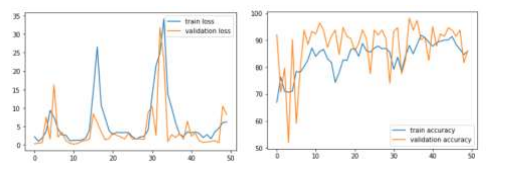}
  \caption{Learning curves for DenseNet121.}
  \label{fig:learn_densenet}
\end{figure}

\begin{figure}[H]
  \centering
  \includegraphics[width=\linewidth]{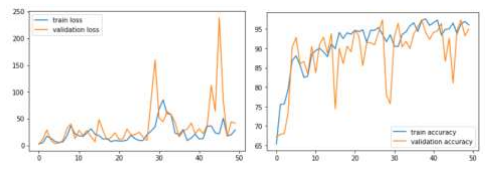}
  \caption{Learning curves for ResNet152.}
  \label{fig:learn_resnet}
\end{figure}

\begin{figure}[H]
  \centering
  \includegraphics[width=\linewidth]{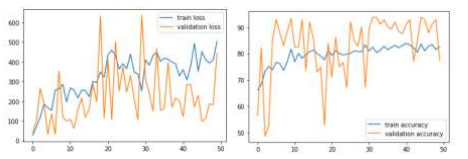}
  \caption{Learning curves for VGG19.}
  \label{fig:learn_vgg}
\end{figure}

Figures~\ref{fig:learn_custom}–\ref{fig:learn_vgg} present the training and validation
accuracy/loss curves for each model, providing insight into convergence behavior, stability, and potential overfitting.

The \textbf{Custom CNN} (Figure~\ref{fig:learn_custom}) achieved steady improvement in both training and validation accuracy over the first 20 epochs, plateauing thereafter at high performance without significant divergence between training and validation curves. This pattern suggests good generalization and minimal overfitting. The loss curves confirm smooth convergence, with validation loss tracking training loss closely.

\textbf{DenseNet121} (Figure~\ref{fig:learn_densenet}) exhibited a rapid accuracy gain within the first 10 epochs, followed by a slower, steady climb. Validation accuracy remained close to training accuracy throughout, and the loss curves showed consistent downward trends for both, indicating strong feature transfer from ImageNet pretraining and effective fine-tuning.

For \textbf{ResNet152} (Figure~\ref{fig:learn_resnet}), training accuracy reached near-perfect levels early, but validation accuracy was more variable, especially in later epochs. The validation loss curve showed small oscillations rather than a monotonic decrease, possibly reflecting sensitivity to hyperparameters or the high capacity of ResNet152, which can overfit when data is limited.

\textbf{VGG19} (Figure~\ref{fig:learn_vgg}) underperformed relative to the other models. While training accuracy improved steadily, validation accuracy plateaued early and even declined in later epochs, accompanied by unstable validation loss. This divergence points to overfitting and insufficient generalization, likely due to the deeper architecture’s reliance on more data for optimal tuning.

\subsection{Quantitative Performance}
Table~\ref{tab:accuracy} reports overall test set accuracy. While \textbf{ResNet152} achieved the highest raw accuracy (\SI{97.3}{\percent}), Table~\ref{tab:metrics} highlights that DenseNet121 outperformed in balanced metrics, achieving the highest macro-averaged precision (\SI{92}{\percent}), recall (\SI{90}{\percent}), and F1-score (\SI{91}{\percent}).

Custom CNN reached moderate accuracy (\SI{92.86}{\percent}) but lower recall than DenseNet121, indicating missed minority-class cases. ResNet152’s strong accuracy was paired with less balanced per-class performance, suggesting dominance by majority-class predictions. VGG19 scored lowest across all metrics, aligning with learning curve findings.
(Table~\ref{tab:metrics}).

\begin{table}[h]
  \centering
  \caption{Overall test accuracy.}
  \label{tab:accuracy}
  \begin{tabular}{@{}lr@{}}
    \toprule
    Model & Accuracy (\%) \\
    \midrule
    Custom CNN & 92.86 \\
    DenseNet121 (fine-tuned) & 89.15 \\
    ResNet152 (fine-tuned) & 97.30 \\
    VGG19 (fine-tuned) & 79.46 \\
    \bottomrule
  \end{tabular}
\end{table}

\begin{table}[h]
  \centering
  \caption{Macro Precision/Recall/F1 on test set.}
  \label{tab:metrics}
  \begin{tabular}{@{}lccc@{}}
    \toprule
    Model & Precision (\%) & Recall (\%) & F1 (\%) \\
    \midrule
    Custom CNN & 78 & 71 & 73 \\
    DenseNet121 (fine-tuned) & 92 & 90 & 91 \\
    ResNet152 (fine-tuned) & 86 & 39 & 42 \\
    VGG19 (fine-tuned) & 85 & 35 & 38 \\
    \bottomrule
  \end{tabular}
\end{table}

\FloatBarrier
\subsection{Confusion Matrices and ROC--AUC}
\begin{figure}[H]
  \centering
  \begin{subfigure}[b]{0.48\linewidth}
    \includegraphics[width=\linewidth]{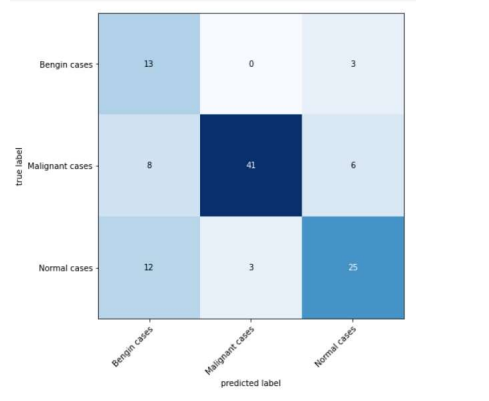}
    \caption{Custom CNN}
  \end{subfigure}\hfill
  \begin{subfigure}[b]{0.48\linewidth}
    \includegraphics[width=\linewidth]{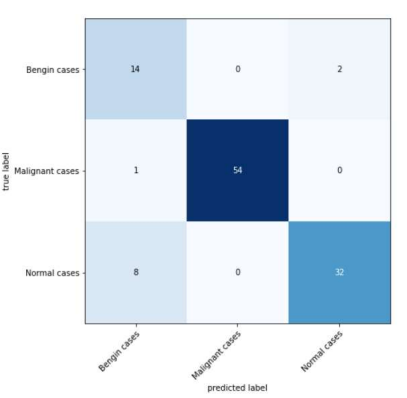}
    \caption{DenseNet121}
  \end{subfigure}\\[6pt]
  \begin{subfigure}[b]{0.48\linewidth}
    \includegraphics[width=\linewidth]{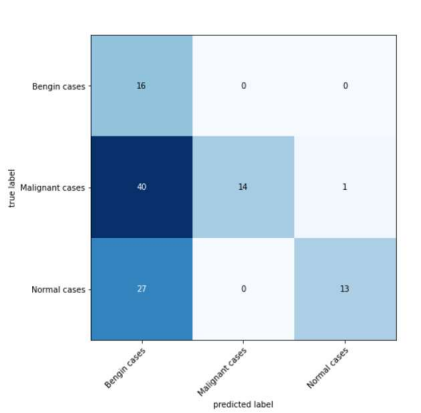}
    \caption{ResNet152}
  \end{subfigure}\hfill
  \begin{subfigure}[b]{0.48\linewidth}
    \includegraphics[width=\linewidth]{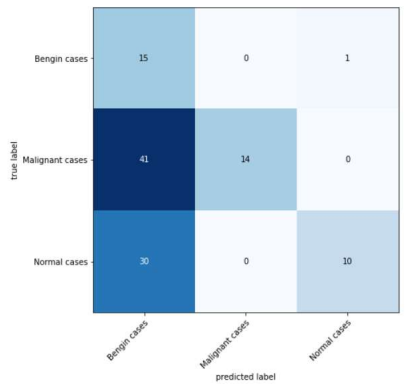}
    \caption{VGG19}
  \end{subfigure}
  \caption{Confusion matrices on the test split.}
  \label{fig:cms}
\end{figure}

Figure~\ref{fig:cms} shows confusion matrices for all models. DenseNet121 demonstrated balanced classification across all three categories, with minimal confusion between Benign and Malignant. Custom CNN effectively identified Normal and Malignant cases but struggled more with Benign vs. Malignant separation. ResNet152 classified most cases correctly but occasionally mislabeled Benign as Malignant—potentially problematic in a clinical context. VGG19 had the most off-diagonal errors.

\begin{figure}[H]
  \centering
  \begin{subfigure}[b]{0.48\linewidth}
    \includegraphics[width=\linewidth]{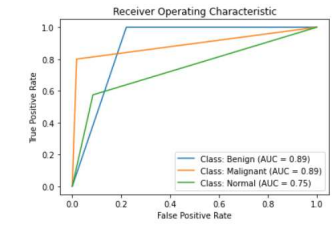}
    \caption{Custom CNN}
  \end{subfigure}\hfill
  \begin{subfigure}[b]{0.48\linewidth}
    \includegraphics[width=\linewidth]{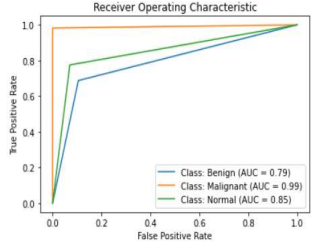}
    \caption{DenseNet121}
  \end{subfigure}
  \caption{ROC--AUC curves per class for Custom CNN and DenseNet121.}
  \label{fig:roc}
\end{figure}

Figure~\ref{fig:roc} displays per-class ROC curves for Custom CNN and DenseNet121. DenseNet121 achieved especially high AUC values (Malignant $\approx 0.99$), underscoring its strong sensitivity to cancerous lesions—crucial in screening where false negatives are costly. Custom CNN also achieved high AUCs but trailed slightly, especially for Benign cases.

\FloatBarrier
\subsection{Explainability with SHAP}
\begin{figure}[H]
  \centering
  \begin{subfigure}[b]{0.32\linewidth}
    \includegraphics[width=\linewidth]{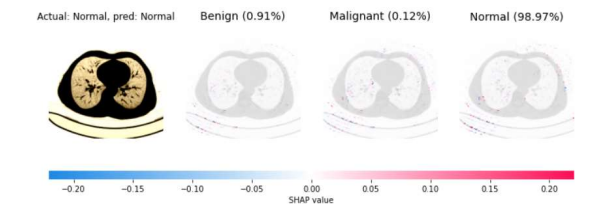}
    \caption{Normal}
  \end{subfigure}
  \begin{subfigure}[b]{0.32\linewidth}
    \includegraphics[width=\linewidth]{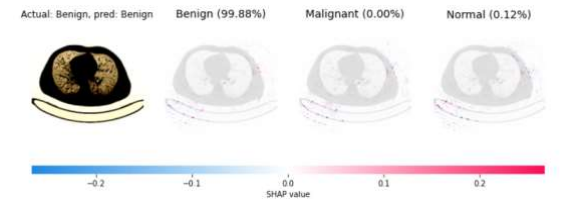}
    \caption{Benign}
  \end{subfigure}
  \begin{subfigure}[b]{0.32\linewidth}
    \includegraphics[width=\linewidth]{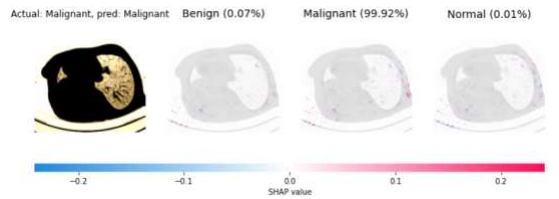}
    \caption{Malignant}
  \end{subfigure}
  \caption{SHAP attributions highlighting evidence for each predicted class.}
  \label{fig:shap}
\end{figure}

Figure~\ref{fig:shap} presents SHAP heatmaps for Normal, Benign, and Malignant examples. Pink highlights indicate features contributing positively to predictions; blue indicates features lowering prediction confidence.

For Malignant cases, SHAP emphasized irregular, dense tissue and spiculated margins—consistent with radiological indicators of malignancy. Benign predictions focused on smooth, well-defined nodules, while Normal cases had diffuse low activations with minor vessel emphasis. DenseNet121’s attention consistently aligned with plausible lung regions, while weaker models like VGG19 occasionally fixated on irrelevant peripheral areas, hinting at overfitting to spurious patterns.

\section{Discussion}
Although ResNet152 reached the highest accuracy, DenseNet121 produced the most balanced precision/recall/F1 and the highest malignant AUC, which can be preferable in clinical screening to minimize false negatives. VGG19 underperformance suggests that deeper VGG stacks are less robust here without extensive tuning or larger training sets. SHAP visualizations provided clinically plausible evidence maps, potentially aiding radiologist review and trust.

\paragraph{Limitations.}
The dataset is modest in size and originates from a single source, which may limit generalization across scanners and institutions. The models operate on 2D slices; incorporating 3D context and volumetric aggregation may improve performance. Finally, SHAP is a post-hoc explanation and does not guarantee causal localization.

\paragraph{Future Work.}
We plan to (i) extend to multi-center datasets, (ii) evaluate 3D CNNs and vision transformers, (iii) integrate cost-sensitive/focal losses and advanced sampling, (iv) apply federated learning for privacy-preserving training, and (v) compare multiple XAI methods (e.g., Integrated Gradients, Grad-CAM++) with user studies.

\section{Conclusion}
We presented an explainable CNN-based pipeline for lung cancer detection on CT images. Across custom and transfer learning models, DenseNet121 offered the best balance of clinical metrics, and SHAP visualizations enhanced interpretability. The approach is promising for decision support, especially where radiology expertise is scarce.

\section*{Acknowledgements}
We would like to express our sincere gratitude to our project supervisor, Er.\ Bikal Adhikari, for his valuable time, guidance, and support throughout this work. 
We also thank our project coordinator, Er.\ Saban Kumar K.C, for his constant encouragement, and Er.\ Chaitya Shova Shakya, DHOD of the Department of Computer and Electronics, Communication, and Information at Kathford International College of Engineering and Management, for providing the opportunity and environment to carry out this project.
We gratefully acknowledge the creators of the IQ{-}OTH/NCCD dataset for making their data publicly available, which was essential to this research. 
Finally, we appreciate the support of our friends, families, and all who contributed directly or indirectly to the successful completion of this work.

\bibliographystyle{IEEEtran}
\bibliography{main}

\begin{thebibliography}{10}
\providecommand{\url}[1]{#1}
\csname url@samestyle\endcsname
\providecommand{\newblock}{\relax}
\providecommand{\bibinfo}[2]{#2}
\providecommand{\BIBentrySTDinterwordspacing}{\spaceskip=0pt\relax}
\providecommand{\BIBentryALTinterwordstretchfactor}{4}
\providecommand{\BIBentryALTinterwordspacing}{\spaceskip=\fontdimen2\font plus
\BIBentryALTinterwordstretchfactor\fontdimen3\font minus \fontdimen4\font\relax}
\providecommand{\BIBforeignlanguage}[2]{{%
\expandafter\ifx\csname l@#1\endcsname\relax
\typeout{** WARNING: IEEEtran.bst: No hyphenation pattern has been}%
\typeout{** loaded for the language `#1'. Using the pattern for}%
\typeout{** the default language instead.}%
\else
\language=\csname l@#1\endcsname
\fi
#2}}
\providecommand{\BIBdecl}{\relax}
\BIBdecl

\bibitem{valente2016svm}
I.~R. Valente \emph{et~al.}, ``A novel svm-based cad system for early detection of lung nodules in computed tomography,'' \emph{Computers in Biology and Medicine}, vol.~69, pp. 157--167, 2016.

\bibitem{gonzalez2016rf}
G.~Gonzalez \emph{et~al.}, ``Automatic lung nodule detection combining rule-based and machine learning approaches,'' \emph{Expert Systems with Applications}, vol.~61, pp. 1--9, 2016.

\bibitem{krizhevsky2012alexnet}
A.~Krizhevsky, I.~Sutskever, and G.~E. Hinton, ``Imagenet classification with deep convolutional neural networks,'' in \emph{Advances in Neural Information Processing Systems}, vol.~25, 2012.

\bibitem{simonyan2015vgg}
K.~Simonyan and A.~Zisserman, ``Very deep convolutional networks for large-scale image recognition,'' \emph{International Conference on Learning Representations}, 2015.

\bibitem{he2016resnet}
K.~He, X.~Zhang, S.~Ren, and J.~Sun, ``Deep residual learning for image recognition,'' in \emph{Proceedings of the IEEE conference on computer vision and pattern recognition}, 2016, pp. 770--778.

\bibitem{huang2017densenet}
G.~Huang, Z.~Liu, L.~Van Der~Maaten, and K.~Q. Weinberger, ``Densely connected convolutional networks,'' in \emph{Proceedings of the IEEE conference on computer vision and pattern recognition}, 2017, pp. 4700--4708.

\bibitem{shen2017multicnn}
W.~Shen, M.~Zhou, F.~Yang \emph{et~al.}, ``Multi-crop convolutional neural networks for lung nodule malignancy suspiciousness classification,'' \emph{Pattern Recognition}, vol.~61, pp. 663--673, 2017.

\bibitem{zhu20183dcnn}
W.~Zhu, C.~Liu, W.~Fan, and X.~Xie, ``Deeplung: 3d deep convolutional nets for automated pulmonary nodule detection and classification,'' in \emph{IEEE Winter Conference on Applications of Computer Vision (WACV)}, 2018.

\bibitem{tajbakhsh2016transfer}
N.~Tajbakhsh \emph{et~al.}, ``Convolutional neural networks for medical image analysis: Full training or fine tuning?'' \emph{IEEE transactions on medical imaging}, vol.~35, no.~5, pp. 1299--1312, 2016.

\bibitem{armato2011lidc}
S.~G. Armato \emph{et~al.}, ``The lung image database consortium (lidc) and image database resource initiative (idri): A completed reference database of lung nodules on ct scans,'' \emph{Medical physics}, vol.~38, no.~2, pp. 915--931, 2011.

\bibitem{wang2017residualattention}
F.~Wang \emph{et~al.}, ``Residual attention network for image classification,'' in \emph{Proceedings of the IEEE Conference on Computer Vision and Pattern Recognition}, 2017, pp. 3156--3164.

\bibitem{dosovitskiy2020vit}
A.~Dosovitskiy \emph{et~al.}, ``An image is worth 16x16 words: Transformers for image recognition at scale,'' in \emph{International Conference on Learning Representations}, 2020.

\bibitem{shen2015multiscale}
W.~Shen, M.~Zhou, F.~Yang \emph{et~al.}, ``Multi-scale convolutional neural networks for lung nodule classification,'' \emph{Information Processing in Medical Imaging}, vol. 9123, pp. 588--599, 2015.

\bibitem{selvaraju2017gradcam}
R.~R. Selvaraju \emph{et~al.}, ``Grad-cam: Visual explanations from deep networks via gradient-based localization,'' in \emph{Proceedings of the IEEE International Conference on Computer Vision}, 2017, pp. 618--626.

\bibitem{sundararajan2017integrated}
M.~Sundararajan, A.~Taly, and Q.~Yan, ``Axiomatic attribution for deep networks,'' in \emph{Proceedings of the 34th International Conference on Machine Learning}, vol.~70, 2017, pp. 3319--3328.

\bibitem{ribeiro2016lime}
M.~T. Ribeiro, S.~Singh, and C.~Guestrin, ``"why should i trust you?": Explaining the predictions of any classifier,'' in \emph{Proceedings of the 22nd ACM SIGKDD International Conference on Knowledge Discovery and Data Mining}, 2016, pp. 1135--1144.

\bibitem{lundberg2017shap}
S.~M. Lundberg and S.-I. Lee, ``A unified approach to interpreting model predictions,'' \emph{Advances in Neural Information Processing Systems}, vol.~30, 2017.

\bibitem{holzinger2019xai}
A.~Holzinger \emph{et~al.}, ``Causability and explainability of artificial intelligence in medicine,'' \emph{Wiley Interdisciplinary Reviews: Data Mining and Knowledge Discovery}, vol.~9, no.~4, p. e1312, 2019.

\bibitem{iqoth_dataset}
{Iraqi Oncological Teaching Hospital (IQ{-}OTH) and Iraqi Center for Cancer and Medical Genetic Research (NCCD)}, ``{IQ{-}OTH/NCCD Lung Cancer CT Scan Dataset},'' \url{https://www.kaggle.com/datasets/hamdallak/the-iqothnccd-lung-cancer-dataset}, 2019, accessed: 2025-08-13.

\end{thebibliography}

\end{document}